\documentclass[prb,twocolumn,showpacs,floatfix]{revtex4}
\usepackage{epsfig,subfigure,amssymb,amscd,amsmath,graphicx,amsfonts,wasysym}
\usepackage{times}

\newcommand{\nin}{\noindent}
\newcommand{\non}{\nonumber}

\newcommand{\bea}{\begin{eqnarray}}
\newcommand{\eea}{\end{eqnarray}}
\newcommand{\be}{\begin{equation}}
\newcommand{\ee}{\end{equation}}
\newcommand{\ba}{\begin{align}}
\newcommand{\ea}{\end{align}}

\newcommand{\ket}[1]{     |    \,    #1    \rangle}
\newcommand{\bra}[1]{  \langle #1  \,  |} 
\newcommand{\ZZ}{\mathbb{Z}}

\newcommand{\anb}{b^{\phantom\dagger}}
\newcommand{\crb}{b^\dagger}
\newcommand{\anc}{c^{\phantom\dagger}}
\newcommand{\crc}{c^\dagger}

\newcommand{\bk}{{\boldsymbol{k}}}

\newcommand{\bq}{{\boldsymbol{q}}}

\newcommand{\bn}{{\boldsymbol{n}}}
\newcommand{\bs}[1]{ \boldsymbol{#1} }
\begin{document}

\title{Exact results for the star lattice chiral spin liquid}

\author{G. Kells$^{1}$, D. Mehta$^{1}$, J. K. Slingerland$^{1,2}$ and J. Vala$^{1,2}$}

\affiliation{$^{1}$  Department   of  Mathematical  Physics,  National University of Ireland, Maynooth, Ireland, \\ $^{2}$ Dublin Institute for Advanced  Studies, School of Theoretical  Physics, 10 Burlington Rd, Dublin, Ireland. }

\begin{abstract}
We examine the star lattice Kitaev model whose ground state is a a chiral spin liquid. We fermionize the model such that the fermionic vacua are toric code states on an effective Kagome lattice. This implies that the Abelian phase of the system is inherited from the fermionic vacua and that time reversal symmetry is spontaneously broken at the level of the vacuum. In terms of these fermions we derive the Bloch-matrix Hamiltonians for the vortex free sector and its time reversed counterpart and illuminate the relationships between the sectors. The phase diagram for the model is shown to be a sphere in the space of coupling parameters around the triangles of the lattices. The Abelian phase lies inside the sphere and the critical boundary between topologically distinct Abelian and non-Abelian phases lies on the surface. Outside the sphere the system is generically gapped except in the planes where the coupling parameters between the vertices on triangles are zero. These cases correspond to bipartite lattice structures and the dispersion relations are similar to that of the original Kitaev honeycomb model. In a further analysis we demonstrate the three-fold non-Abelian groundstate degeneracy on a torus by explicit calculation.
\end{abstract}


\date{\today}
\maketitle

\section{Introduction}

In his original analysis of the honeycomb model \cite{Kitaev06}, Kitaev noted that a similar type of system but with triangles placed at the vertices of the honeycomb lattice would spontaneously break time reversal symmetry.  A system of precisely this type was subsequently analyzed  by Yao and Kivelson \cite{YaoKivelson07} and shown to be an example of a chiral spin liquid with just nearest neighbor interactions between sites. This system also inherits a number of interesting properties from the original honeycomb model, particularly the existence of both Abelian and non-Abelian topological phases. A finite temperature analysis of the model has recently been performed  \cite{Chung10} and, as with the original honeycomb model, there are important overlaps with the physics of classical dimer models and Kasteleyn matrices \cite{Nash09}.  

In what follows we present an analysis of this system using the Jordan-Wigner fermionization procedure of Ref.~\onlinecite{Kells09b} which explicitly formulates the fermionic vacua as toric code states on an effective Kagome lattice. The fermionization procedure is a two step process where we map the model to a system of hard-core bosons and spins on an effective Kagome lattice \cite{Dusuel08b,Schmidt07}, and then define fermions in terms of the hard-core bosons and spins. Once fermionized, and similarly to the original honeycomb lattice model \cite{Chen07}, the ground-state sector of the system can be transformed to that of a spinless p-wave superconductor.  With our method however, we obtain vacua for the fermionized theory as the stabilized wavefunctions of an abelian toric code model \cite{Kitaev03}, defined with effective spins on a Kagome lattice. These vacuum wavefunctions are independent of the couplings of the model. The ground-state for the full system, valid for the entire parameter space, is a BCS type condensate over the toric code ground-state. The topological degeneracies of the model are already present at the level of the fermionic vacuum. However, by generalizing the arguments presented in \cite{Kells09b,ReadGreen00} we show how the BCS product lifts some of this degeneracy in the non-Abelian phase. The predicted degeneracy is in agreement with the original observations of Yao and Kivelson \cite{YaoKivelson07}.  

The representation we use also illuminates the nature of the spontaneously broken time-reversal symmetry. We see for example that this symmetry is broken at the level of the vacua. The chiral nature of these vacuum states has been recently analyzed \cite{WangWan08}.  A detailed analysis of the phase boundary between the Abelian and the non-Abelian phase is also included and we show that this critical boundary can be understood as a sphere in the space of the spin-spin coupling strengths. We also see that the opening of a gap to a non-Abelian phase is highly dependent on the interaction strengths around the triangular plaquettes. We will show that if any of these interactions are subtracted from the Hamiltonian we cannot open a gap to the non-abelian phase.  
Exact expressions for the energy dispersion relations for ground states sectors are proved in the appendix, which also provides a brief analysis of the Dirac cone structures of the gapless phases.

\section{The star lattice Kitaev model and the hexagonal toric code}

\begin{figure}
\includegraphics[width=.4\textwidth,height=0.3\textwidth]{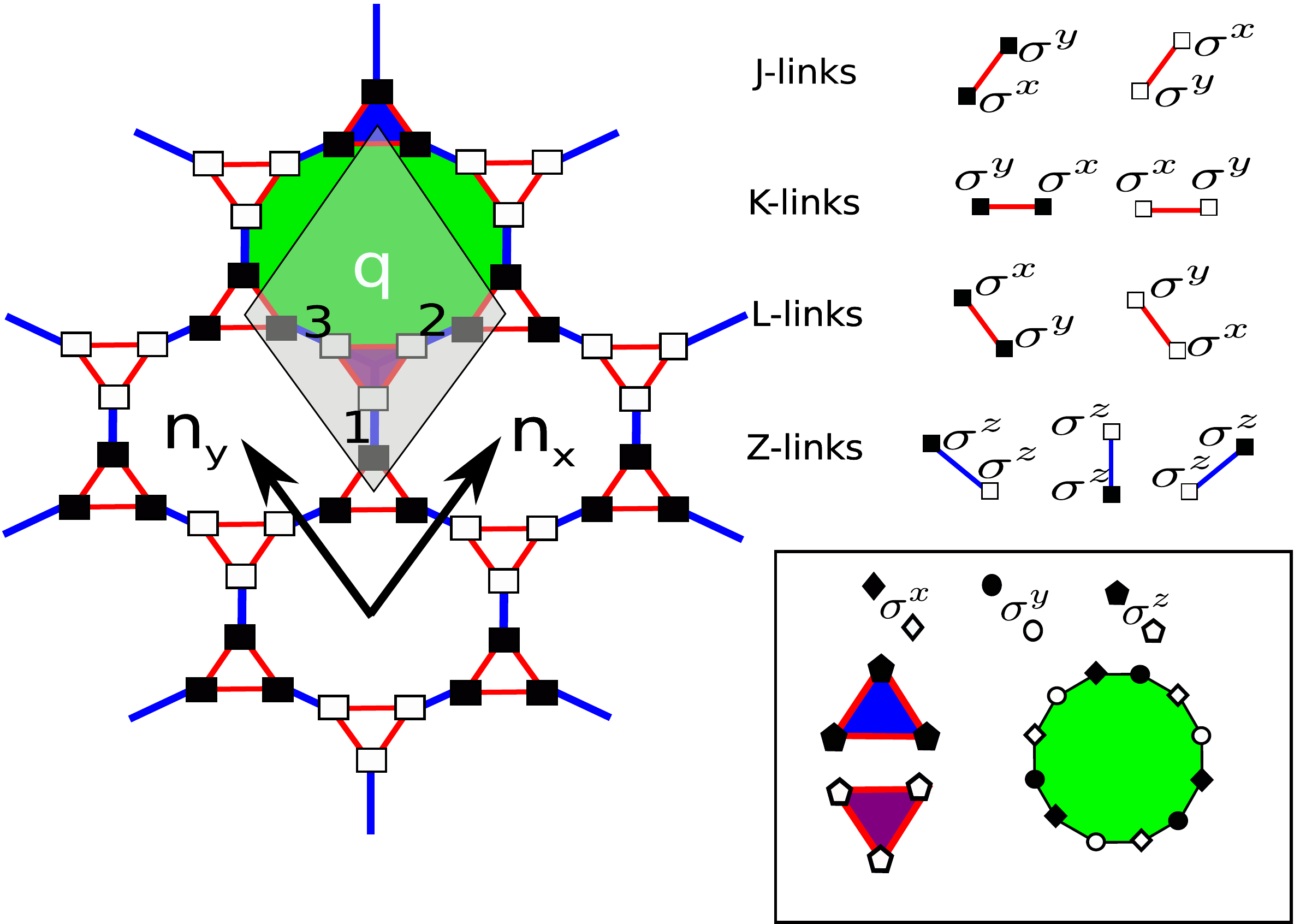}
       \caption{(Color online) The star lattice is a hexagonal lattice with each vertex replaced by a triangle. There are two types of plaquettes symmetries in the system: triangular and dodecagonal. The triangular symmetries are responsible for the spontaneous breaking of time-reversal symmetry \cite{Kitaev06,YaoKivelson07}}
       \label{fig:HTlattice}
\end{figure}

The Hamiltonian consists of directional spin-spin interactions on the star lattice (also known as the triangle-honeycomb or Fisher lattice).  We use the representation of the model introduced in Ref.~\onlinecite{Dusuel08b} as it provides a straightforward route , by contracting the $Z$-links, to the definition of the fermions as toric code states on the Kagome lattice. In this representation the Hamiltonian can be written as 
\bea 
\label{eq:H}
H &=& H_Z+H_J+H_K +H_L\\ &=& - Z \sum_{\text{Z-links}}  \sigma^z \sigma^z  - J
\sum_{ \text{J-links} } \sigma^x \sigma^y \non \\  && ~~~~~~~ - K \sum_{\text{K-links}} \sigma^x \sigma^y - L \sum_{\text{L-links}} \sigma^x \sigma^y \non
\eea 
where it should be understood that the $Z$-links connect separate triangles and the $J$,$K$ and $L$-links within the triangles are the positive, zero and negative slopes respectively, see Figure \ref{fig:HTlattice}. We refer to triangles that point up as black triangles and those that point down as white triangles. The sites on these triangles in the original lattice are colored black and white respectively. 

We define a basic unit cell of the lattice around a white triangle. We label the $Z$-link at the bottom of the triangle with $n=1$, the $Z$-link from the top right with $n=2$ and the $Z$-link from the top left with  $n=3$. Each spin site can be specified using the position vector $\bq$, the index $n$, and whether it is on a $\square$ or a $\blacksquare$ site. In a $6N$ spin system we have $N$ unit cells.

\begin{figure}
\includegraphics[width=.3\textwidth,height=0.25\textwidth]{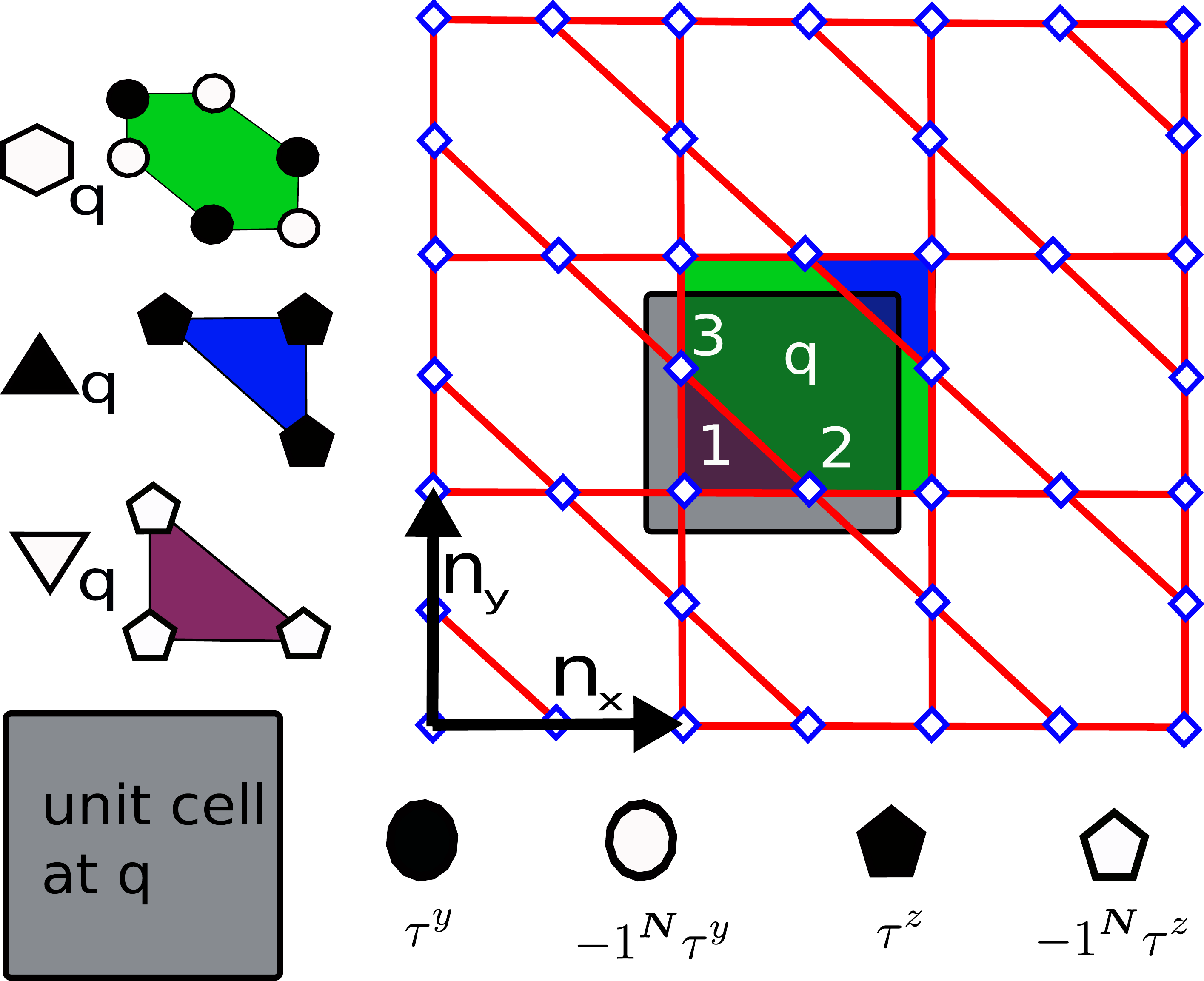}
       \caption{(Color online) The effective Kagome lattice and plaquette operators. Note that within our labeling convention the ${{\blacktriangle}_\bq} $ operator does not actually contain any sites from the unit cell at $\bq$.}
       \label{fig:lattice2}
\end{figure}

Using this basic unit cell we we may write 
\be
H_Z= -Z \sum_\bq \sum_{n=1}^3 \sigma^z_{\bq,n,_\blacksquare } \sigma^z_{\bq,n,_\square},
\ee 
\be
H_J =  -J {\sum} _\bq  \sigma^y_{\bq,1,_\square } \sigma^x_{\bq,2,_\square} + \sigma^y_{\bq\rightarrow,1,_\blacksquare} \sigma^x_{\bq,2,_\blacksquare }   ,
\ee
\be
H_K = -K {\sum}  _\bq  \sigma^y_{\bq,2,_\square } \sigma^x_{\bq,3,_\square}  + \sigma^y_{\bq\nwarrow,2,_\blacksquare} \sigma^x_{\bq,3,_\blacksquare } , 
\ee
\be
H_L = -L  {\sum} _\bq \sigma^y_{\bq,3,_\square } \sigma^x_{\bq,1,_\square} 
 + \sigma^y_{\bq\downarrow,3,_\blacksquare }  \sigma^x_{\bq,1,_\blacksquare}  ,
\ee 
where have introduced here the shorthand notation $\small{\bq \nwarrow = \bq-\bs{n}_x+\bs{n}_y}$, $\small{\bq \downarrow = \bq-\bs{n}_y}$ etc. and the two unit vectors $\bs{n}_x$ and $\bs{n}_y$ as shown in FIG. \ref{fig:HTlattice}.  We have set all coupling strengths on all $z$-dimers to $Z$ thus restricting the parameter space of the original model in this direction. 

Within the model, as in honeycomb lattice model, there are closed loop symmetries that we can generate (up to a phase) with overlapping link operators. The simplest of these are the dodecagonal and triangular loops. These are defined pictorially in Figure \ref{fig:HTlattice}. For simplicity we will refer to a generic loop symmetry as $\bs{W}_\bq$ and define it such that it has eigenvalues of $\pm 1$. The fact that the Hamiltonian commutes with all plaquette operators implies that we may choose energy eigenvectors $\ket{n}$ such that $W_{\bq}=\bra{n} \bs{W}_\bq \ket{n}=\pm 1$. If $W_{\bq} = -1$ then we say that the state $\ket{n}$ carries either a triangular or dodecagonal vortex at $\bq$. When we refer to a  vortex-sector we mean the subspace of the system Hilbert space with a particular configuration of vortices.  The vortex-free sector for example is the subspace spanned by all eigenvectors such that $W_{\bq} = 1$ for all $\bq$. We will have cause later to distinguish between triangular and dodecagonal plaquettes. Our labeling convention will however reflect the Kagome lattice on which the fermions (and hard-core bosons and effective spins) are defined. 

On a torus, the plaquette operators are not independent, as they obey $\prod W_{\bq} =I$ where the product is over all $\bq$. There are also two independent homologically non-trivial loop symmetries. To represent these we are free to choose any two closed loop operators that traverse the torus as long as they cannot be deformed into each other by plaquette multiplication. All other homologically non-trivial loop symmetries can be constructed from the products of these two operators and the $3N-1$ independent plaquette operators. When the torus is specified by periodic boundary vectors $(\bs{x}, \bs{y})$ which are integer multiples of the unit vectors   i.e. $\bs{x} = N_x \bs{n}_x$ and $\bs{y} = N_y \bs{n}_y$, it is natural to define operators $L^{(x)}_{q_y}=\prod_{q_x} \sigma^x_{\bq,1,_\square} \sigma^x_{\bq,1,_\blacksquare}  \sigma^y_{\bq,2,_\square} \sigma^y_{\bq,2,_\blacksquare}$ and $L^{(y)}_{q_x}=\prod_{q_y} \sigma^y_{\bq,1,_\square} \sigma^y_{\bq,1,_\blacksquare}  \sigma^x_{\bq,3,_\square} \sigma^x_{\bq,3,_\blacksquare}$ as the homologically non-trivial symmetries. We will generally use the operators $L^{(x)}_{0}$ and $L^{(y)}_{0}$ that run through the origin as the two independent symmetries.
For an analysis of the loop symmetries in the original honeycomb model see Ref.~\onlinecite{Kells08a}.

It was pointed out by Kitaev that a model of this type (i.e with triangles at the vertices of a honeycomb lattice) must spontaneously break time reversal symmetry. This implies that all states of the system are at least two fold degenerate \cite{Kitaev06}. To see this let $T$ be the time reversal operator. Now, because $T \sigma^\alpha = -\sigma^\alpha$, the time reversal operator changes the eigenvalues of all triangular plaquette operators. However, the operator itself commutes with the Hamiltonian as it contains only terms of the form $\sigma^a \sigma^b$. Each eigenstate must therefore have a time-reversed counterpart with the same energy, but from the vortex sector where the eigenvalues of all triangular plaquettes are negated \cite{TRsym}. 

The Hamiltonian (\ref{eq:H}) can be written in terms of hard-core bosons and effective spins of the $Z$-dimers using the mapping \cite{Dusuel08b,Schmidt07}:
\bea 
\label{eq:map}
\ket{  \uparrow_{_\blacksquare} \uparrow_{_\square}}  &=& \ket{\Uparrow,0},
\quad
\ket{\downarrow_{_\blacksquare} \downarrow_{_\square}} = \ket{\Downarrow,0},
\\
\ket{ \uparrow_{_\blacksquare}  \downarrow_{_\square}} &=& \ket{\Uparrow,1},
\quad
\ket{  \downarrow_{_\blacksquare} \uparrow_{_\square}} = \ket{\Downarrow,1}.
\non
\eea

The labels on the left hand side indicate the states of the z-dimer in the $S_z$ basis. The first quantum number of the kets on the right hand side represents the effective spin of the square lattice and the second is the hard-core bosonic occupation number. The presence of a boson indicates an anti-ferromagnetic configuration of the spins connected by a $z$-link.  

In the Abelian phase, the dominance of the $Z$-coupling terms means that spins on these links tend to align. In this limit the bosons are energetically suppressed. A perturbative analysis for the low energy effective Hamiltonian in this regime shows that the first non-constant terms occur at the 6th order and 8th orders. We have from Ref.~\onlinecite{Dusuel08b}
\be
H_{\text{eff}} = -J_{\text{eff}}^{(6)} \sum_\bq {{\varhexagon}}_{\bq}  - J_{\text{eff}}^{(8)} \sum_\bq P( {\large{\blacktriangle}}_{\bq_i} {{\varhexagon}}_{\bq_j}  \triangledown_{\bq_k} )
\label{eq:HE}
\ee 
with ${{\varhexagon}}_{\bq}= \prod \tau_{\bs{q}}^y$, ${\bs{\triangledown}}_{\bq}= \prod_{_\square} \tau_{\bs{q}}^z$ and ${\large{\blacktriangle}}_{\bq}= \prod_{_\blacksquare} \tau_{\bs{q}}^z$  where $\tau_\bq^a$ is the Pauli operator acting on the effective spin at position $\bq$. The functional $P$ refers to combinations of a hexagon with two attached triangles.  This effective Hamiltonian, defined now on a Kagome lattice, is unitarily equivalent to what is known as the hexagonal toric code, see for example  Ref.~\onlinecite{WangWan08}. The hexagonal toric code shares many of the same properties as the original square toric code system of Kitaev \cite{Kitaev03}. All eigenstates of the effective Hamiltonian (\ref{eq:HE}) on a plane may be completely characterized by the set of eigenvalues $\{ \varhexagon_\bq ,\triangledown_{\bq}, \blacktriangle_{\bq} \}$ for all $\bq$. The ground states are those with all plaquette eigenvalues equal to +1 and its time reversed counterpart with all triangular plaquettes equal to -1 and on all hexagonal plaquette eigenvalues equal to +1 \cite{YaoKivelson07,Dusuel08b}. On a torus of $N$ unit cells with $3N$ effective spins we have the following identities:
\be
\prod_\bq^N \blacktriangle_\bq \triangledown_\bq =I,  \quad \prod_\bq^N \varhexagon_\bq =I
\ee
and so we have there a total of $3N-2$ independent plaquette symmetries. However we gain two independent homologically non-trivial symmetries and thus eigenstates on a torus are uniquely labeled by using the full set of independent symmetries. 

The basis (\ref{eq:map}) also describes anti-ferromagnetic configurations of the z-dimers through the bosonic occupation number and forms an orthonormal basis for the full star lattice system.  The Pauli operators of the original spin Hamiltonian can be written as (see Refs.~\onlinecite{Schmidt07}) :
\begin{equation}
    \label{eq:mapping}
    \begin{array}{lcl}
    \sigma_{\bq,_\blacksquare}^x=\tau_\bq^x  (\crb_\bq+ \anb_\bq) &,& 
\sigma_{\bq,_\square}^x=\crb_\bq+\anb_\bq  ,\\  %
    \sigma_{\bq,_\blacksquare}^y=\tau_\bq^y (\crb_\bq+\anb_\bq) &,& 
\sigma_{\bq,_\square}^y= i \,  \tau^z_\bq (\crb_\bq-\anb_\bq), \\ 
\sigma_{\bq,_\blacksquare}^z=
    \tau_\bq^z  &,&  \sigma_{\bq,_\square}^z=\tau_\bq^z (I-2  \crb_\bq
    \anb_\bq),
    \end{array}
\end{equation}
where $\crb$ and $\anb$ are the creation and annihilation operators for the hard-core bosons. In this representation we have
\be
H_Z = -Z \sum_{\bq,n} (I -2 \crb_{\bq,n} \anb_{\bq,n}) ,
\label{eq:HbZ}
\ee
\bea
\label{eq:HbJ}
H_J &=& - J {\sum} _{\bq} [ i  \tau^z_{\bq,1} (\crb_{\bq,1} - \anb_{\bq,1})(  \crb_{\bq,2} + \anb_{\bq,2}) \\ \non  && +  \tau^x_{\bq,2} (\crb_{\bq,2}+\anb_{\bq,2}) \tau^y_{\bq\rightarrow,1} (\crb_{\bq\rightarrow,1}+\anb_{\bq\rightarrow,1}) ]
\eea
\bea
 \label{eq:HbK}
H_K &=& - K {\sum} _{\bq} [  i  \tau^z_{\bq,2} (\crb_{\bq,2} - \anb_{\bq,2})(  \crb_{\bq,3} + \anb_{\bq,3})  \\ \non &+&   \tau^x_{\bq,3} \tau^y_{\bq\nwarrow,2} (\crb_{\bq,3}+\anb_{\bq,3})  (\crb_{\bq\nwarrow,2}+\anb_{\bq\nwarrow,2}) ] 
\eea
\bea
\label{eq:HbL}
H_L &=& -L {\sum} _{\bq} [ i \tau^z_{\bq,3} (\crb_{\bq,3} - \anb_{\bq,3})(  \crb_{\bq,1} + \anb_{\bq,1})   \\  \non &+&  \tau^x_{\bq,1} \tau^y_{\bq\downarrow,3}(\crb_{\bq,1}+\anb_{\bq,1})   (\crb_{\bq\downarrow,3}+\anb_{\bq\downarrow,3}) ] 
\eea
\nin The basic plaquette operators written in this basis are
\bea
{\varhexagon}_{\bq} &=& (-1)^{\crb_{\bq,3} \anb_{\bq,3} +  \crb_{\bq\rightarrow,1} \anb_{\bq\rightarrow,1} + \crb_{\bq\uparrow,2} \anb_{\bq\uparrow,2} } \\ \non && \times \tau_{\bq,3~~}^{y}  \tau_{\bq,2~~}^{y}  \tau_{\bq\rightarrow,1}^{y}  \tau_{\bq\rightarrow,3}^{y} \tau_{\bq~\uparrow,2}^{y}  \tau_{\bq~\uparrow,1}^{y},
\eea
for the hexagons and 
\bea
{\large{\blacktriangle}}_\bq&=&  \tau^z_{\bq,1 \nearrow} \tau^z_{\bq\uparrow,2}  \tau^z_{\bq~\rightarrow,3}   \\ 
{\large{\triangledown}}_\bq&=& \prod_{n=1}^3 (-1)^{\crb_{\bq,n} \anb_{\bq,n}} \tau^z_{\bq,n},
\eea
for the triangles. The unit cell and the plaquettes are shown pictorially in Figure \ref{fig:lattice2}. Note that with this labeling convention the ${\blacktriangle}_\bq$ operator does not contain any sites from within the unit cell at $\bq$. 
\begin{figure}
\includegraphics[width=.42\textwidth,height=0.27\textwidth]{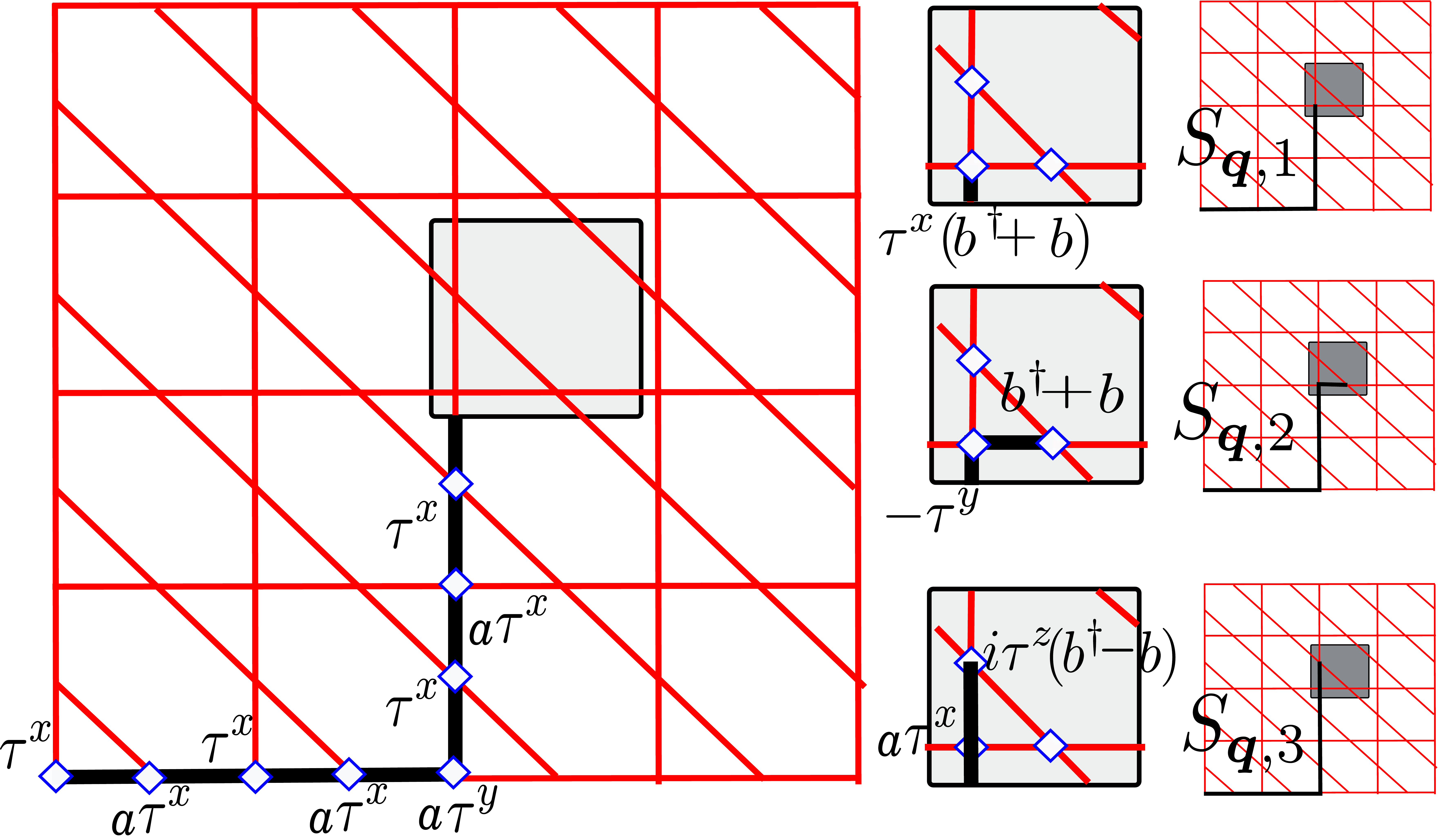}
       \caption{(Color online) The three strings $S_{\bq,1}$, $S_{\bq,2}$, $S_{\bq,3}$ in the effective spin hard-core boson representation. Here $a = (2 b^\dagger b-I)$.}
       \label{fig:S123}
\end{figure}

\begin{figure}
\includegraphics[width=.25\textwidth,height=0.24\textwidth]{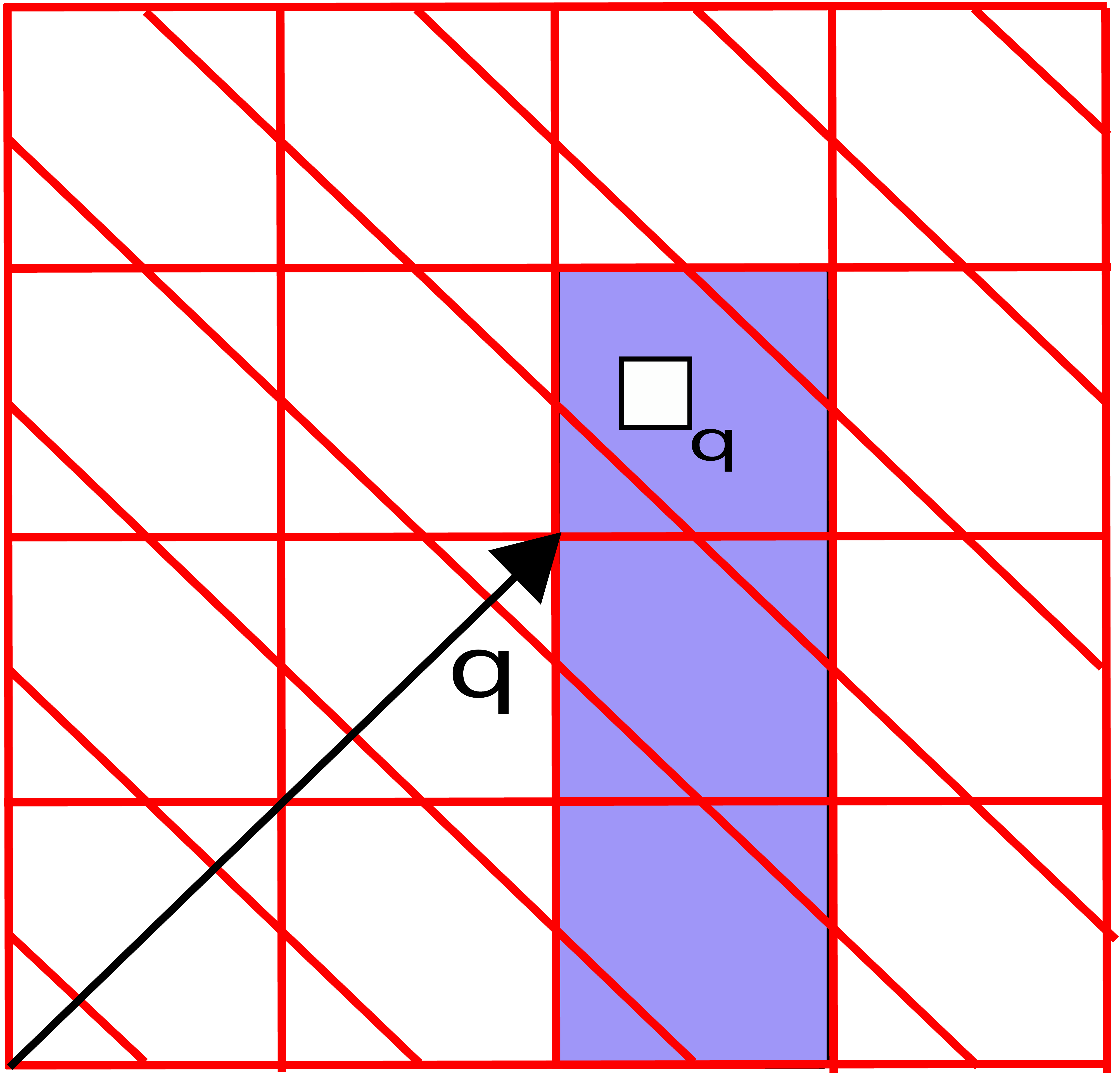}
       \caption{(Color online)The operator $\square_\bq$ and the vector $\bq$. The operator  $\square_\bq$ is the product of all hexagonal and triangular plaquettes in the shaded region}
       \label{fig:squareq}
\end{figure}

\section{Fermionization}
To fermionize the problem we follow the Jordan-Wigner procedure used in Ref.~\onlinecite{Kells09b} for the honeycomb lattice model. To each of the hard-core bosons in the unit cell we attach a string. The strings for hard-core bosons inside the same unit cell are identical everywhere except inside the unit cell, where they branch out. These strings, shown pictorially in Fig \ref{fig:S123}, are designed so that they break/fix a $Z$-dimer at the desired point on the lattice.  This will ensure that the presence of a fermion implies an anti-ferromagnetic configuration of the z-dimer. 

We define the string operators using the following convention: First apply a single $\sigma^y_{\bs{0},1}$ term to a black site of the $Z$-link which we set to be the origin.  The rest of the string is made by operating cyclically with the overlapping links of the Hamiltonian, $\sigma_i^z \sigma_j^z$, $\sigma_j^y \sigma_k^x$,  $\sigma^z_k \sigma^z_l$ and $\sigma^x_l \sigma^y_m$, in the $n_x$ direction until we reach a required length and then $\sigma^z_m \sigma^z_n$, $\sigma_n^x \sigma_o^y$,  $\sigma^z_o \sigma^z_p$ and $\sigma_p^y \sigma_q^x$ in the $\bn_y$ direction until we reach the unit cell at $\bq$. The ends of the string depend on the $Z$-link in question. 

In the original spin representation we can write
\bea
\label{eq:Sq1}
 S_{\bq,n} && \equiv    A_{\bq,n} \times \sigma^x_{(q_x,q_{y-1}),3,_\blacksquare} 
\sigma^x_{(q_x,q_{y-1}),3,_\square} ~~ ... \\
 \non  && \sigma^y_{(q_x,1),1_\blacksquare} \sigma^x_{(q_x,0),3,_\blacksquare}
\sigma^x_{(q_x,0),3,_\square} \sigma^y_{(q_x,0),1,_\square} \sigma^x_{(q_x,0),1,_{\blacksquare}}  \\ \non &&
...  ~ \sigma^x_{(1,0),1,_\blacksquare} \sigma^y_{(0,0),2,_{\blacksquare}} \sigma^y_{(0,0),2,_\square}
\sigma^x_{(0,0),1,_\square} \sigma^x_{(0,0),1,_\blacksquare},
\eea
with 
\bea
A_{\bq,1}&=&\sigma^x_{(q_x,q_y),1,_\blacksquare} \\ \non
A_{\bq,2}&=&\sigma^x_{(q_x,q_y),2,_\square} \sigma^x_{(q_x,q_y),1,_\square} \sigma^y_{(q_x,q_y),1,_\blacksquare}\\ \non
A_{\bq,3}&=&\sigma^y_{(q_x,q_y),3,_\square} \sigma^y_{(q_x,q_y),1,_\square} \sigma^y_{(q_x,q_y),1,_\blacksquare}. 
\eea
The hard-core-boson/effective-spin representation of the strings for each of the three $Z$-links is shown in Figure \ref{fig:S123}. Note that the end of the string always contains a term $(\crb+\anb)$ or $(\crb-\anb)$ thus ensuring a hard-core boson is created or destroyed.  

The operators $S_{\bq,n}$ square to unity while different operators $S_{\bq,m}, S_{\bq',n}$ anti-commute with each other. This leads us to identify the strings $S_{\bq,n}$ with the following sum of fermionic creation and annihilation operators: $S_{\bq,n} = c^\dagger_{\bq,n} + c_{\bq,n} = (\crb_{\bq,n} + \anb_{\bq,n})  S_{\bq,n}^{'}$ where $S^{'}_{\bq,n}$ is simply the string $S_{\bq,n}$ but with the $(\crb_{\bq,n} + \anb_{\bq,n})$ bosonic dependence of the end-point removed,  see Figure \ref{fig:S123}. Individually our fermionic canonical creation and annihilation operators are 
\be
\label{eq:cdef}
\crc_{\bq,n} = \crb_{\bq,n} S^{'}_{\bq,n}, \quad c_{\bq,n} = b_{\bq,n} S^{'}_{\bq,n}
\ee
where the strings now ensure that all operators $c_{\bq,n}^\dagger$ and $c_{\bq,n}$ obey the canonical fermionic anti-commutator relations
\bea
&&\{ c^\dagger_{\bq,m}, c^{\phantom \dagger}_{\bq',m} \} = \delta_{\bq \bq'} \delta_{n,m}, \\ \non && \{ c^{ \dagger}_{\bq,n} ,c^{\dagger}_{\bq',m} \}  =0 ,\quad \{ c^{\phantom \dagger}_{\bq,n} ,c^{\phantom \dagger}_{\bq',m} \} =0 .
\eea

Note that the start of each string $S_\bq$ violates the plaquette symmetries $\blacktriangle_{\bs{0}}$ and $\varhexagon_{\bs{0} \leftarrow}$. This linking of fermions and vortex-pairs seems to be a generic property of Kitaev models that has important consequences for the spectral properties on a torus \cite{Kells09b}

Substituting the $b^\dagger_{\bq,n}=S'_{\bq,n} c^\dagger_{\bq,n}$ and $b^{\phantom \dagger}_{\bq,n}=S'_{\bq,n} c^{\phantom \dagger}_{\bq,n}$ into expressions (\ref{eq:HbZ}), (\ref{eq:HbJ}), (\ref{eq:HbK})  and (\ref{eq:HbL}) gives the following fermionic expressions for the Hamiltonian terms:

\be
H_Z = Z \sum_{\bq,n} (2 \crc_{\bq,n} \anc_{\bq,n} -I)
\label{eq:HfZ}
\ee 
\bea
\label{eq:HfJ}
\non H_J = J {\sum}  _\bq [  (\crc_{\bq~,1} -\anc_{\bq~,1}) (\crc_{\bq~,2} +\anc_{\bq~,2}) \\ + {{\Square}}_{\bq\downarrow} (\crc_{\bq~,2} -\anc_{\bq~,2}) (\crc_{\bq\rightarrow,1} +\anc_{\bq\rightarrow,1})] 
\eea
\bea
\label{eq:HfK}
\non H_K =  K {\sum}  _\bq [\non  - i\triangledown_\bq (\crc_{\bq~,2} +\anc_{\bq~,2})((\crc_{\bq~,3} +\anc_{\bq~,3}) \\ - i {\Square}_{\bq \downarrow} {{\blacktriangle}}_{\bq\downarrow} (\crc_{\bq~,2} -\anc_{\bq~,2})  (\crc_{\bq\searrow,3} -\anc_{\bq \searrow,3}) ]
\eea
\bea
\label{eq:HfL} 
\non H_L =  L {\sum} _\bq [(\crc_{\bq~,1} -\anc_{\bq~,1}) (\crc_{\bq~,3} +\anc_{\bq~,3}) \\
 +  (\crc_{\bq ~,3} -\anc_{\bq ~,3}) (\crc_{\bq \uparrow,1} +\anc_{\bq\uparrow,1}) ].
\eea
where ${\Square}$ is the rectangular product of plaquette operators (hexagons and triangles) shown in the Figure \ref{fig:squareq}.  On a torus the terms that connect opposite sides will have some dependence on the homologically non-trivial loop symmetries. Details on how to calculate their precise values can be found in Ref. \onlinecite{Kells09b}. 

The Jordan-Wigner transform has been chosen so that the vacuum states for the fermions are the toric code states on the effective Kagome lattice.  This immediately implies that that under time reversal the fermionic vacua must be exchanged. The form of the fermionic Hamiltonian also reveals a number of important features. We see that the triangular vorticity is incorporated within the $H_J$ term and the $H_K$ term, through the $\Square_\bq$ operator. However, the eigenvalue of the  $\Square_\bq$ does not change under time reversal as it contains an even number of triangles.  This means that if $K=0$ then eigenstates  $\ket{\psi}$ and $T \ket{\psi}$ are fermionically identical. This would seem like a contradiction ( recall these are different sectors ) but we are saved by recalling that our formulation also demands that the vacuum sectors be defined in terms of the hexagonal toric code on the effective Kagome lattice. We see that fermionically these eigenstates are the same, they have the same fermion density and the same fermion number parity. Indeed they are structurally identical in every way except for the vacuum from which they were created. In the opposite sense, in terms of the fermions at least, any sign of spontaneous symmetry breaking only occurs in the terms in $H_K$, which always contain an odd number of triangular plaquettes. These terms closely resemble the time reversal invariant terms in the honeycomb model \cite{Kitaev06,Kells09b}.  We will see later that, in the same way, these terms are jointly responsible for the opening of a gap.  

To proceed we first re-write the fermonic Hamiltonian as 
\bea
H=  \frac{1}{2} \sum_{\bq \bq'} \left[\begin{array}{cc} c^\dagger_{\bq} & c_\bq
\end{array}  \right] \left[
\begin{array}{cc} \xi_{\bq \bq'} & \Delta_{\bq \bq'} \\ \Delta^\dagger_{\bq
\bq'} & -\xi^{T}_{\bq \bq'} \end{array} \right] \left[\begin{array}{c}
c_{\bq'}
\\ c^\dagger_{\bq'} \end{array}  \right] 
\label{eq:Hg} 
\eea
where the $\bq$ now label both the position and internal indices. The system is diagonalized by solving the Bogoliubov-De Gennes eigenvalue problem
\bea
\left[ \begin{array}{cc} \xi & \Delta \\ \Delta^\dagger & -\xi^{T} \end{array} \right] =  \left[\begin{array}{cc} U & V^* \\ V & U^* \end{array}  \right]  \left[ \begin{array}{cc} E & \bs{0} \\ \bs{0} & -E \end{array}  \right] \left[ \begin{array}{cc} U & V^* \\ V & U^* \end{array}   \right]^\dagger,
\label{eq:BdG}
\eea
where the non-zero entries of the diagonal matrix $E_{nm} = E_n \delta_{nm} $ are the quasiparticle excitation energies. The Bogoliubov-Valentin quasiparticle excitations are 
\bea
\label{eq:gamma}
&& \non \left[\begin{array}{cc}\gamma_1^\dagger,...,\gamma_M^\dagger  & \gamma_1,...,\gamma_M \end{array}  \right] \\&& =  \left[\begin{array}{cc} c_1^\dagger,..., c_M^\dagger & c_1,...,c_M \end{array}  \right]  \mathcal{W} ,
\eea
where 
\be
\mathcal{W} \equiv \left[\begin{array}{cc} U & V^* \\ V & U^* \end{array}\right]  . 
\ee
Inverting (\ref{eq:gamma}) and substituting into (\ref{eq:Hg}) gives
\be
H= \sum_{\bs{n}=1}^{M} E_n (\gamma^\dagger_n \gamma^{\phantom \dagger}_n - \frac{1}{2}) .
\label{eq:Hd}
\ee
Normally one assumes that all the values of $E_n$ in this equation are positive. It is this choice that one usually uses to obtain the ground state energy of $-\sum E_n/2$. However, the Hartree-Fock-Bogoliubov formulation above actually only requires that the values of $E_n$ in (\ref{eq:Hg}) come as negated pairs. It does not specify that positive energies must be associated with $\gamma^\dagger$ operators rather than $\gamma$ operators. This is an important point as physical situations arise naturally where the annihilation of a quasi-particle costs energy.

In the vortex-free sector (eigenvalues of all plaquettes are +1) we can move to momentum space with the Fourier Transform
\be
c_{\bq,n} = M^{-1/2} \sum c_{\bk,n} e^{i \bk \cdot \bq}. 
\ee
After some manipulation we then arrive at the following momentum space representation for the planar Hamiltonian
\be
H=  \frac{1}{2} \sum_{\bk,nm} \left[\begin{array}{cc} c^\dagger_{\bk n} & c_{-\bk n}
\end{array}  \right] H(\bk) \left[\begin{array}{c}
c_{\bk m}
\\ c^\dagger_{-\bk m} \end{array}  \right] 
\label{eq:Hm} 
\ee
with
\be
H(\bk)=\left[\begin{array}{cc} \xi(\bk)  & \Delta(\bk) \\ \Delta(\bk)^\dagger &  -[\xi(-\bk)]^T  \end{array} \right]
\label{eq:Hk}
\ee
where
\be
\xi = \left[ \begin{array}{ccc}  
 2 Z &  J(1+ \theta_x)  &  L(1+\theta_y) \\ 
 J(1+\theta_x^*) & 2 Z  & i K(1+ \theta_y \theta_x^*) \\
 L(1+\theta_y^*)& -i K(1+\theta_x \theta_y^*)  &2 Z  \end{array} \right] ,
\ee
and
\be
\Delta = \left[ \begin{array}{ccc}  
 0 &  J(1- \theta_x) &  L(1-\theta_y) \\ 
 -J(1-\theta_x^*) & 0  & iK (1-\theta_x^* \theta_y) \\
  -L(1-\theta_y^*)& -iK(1-\theta_x\theta_y^*)  &0  \end{array} \right] .
\ee
and we have set $\theta_x =\exp(i k_x)$ and $\theta_y =\exp(i k_y)$.
\begin{figure}
\includegraphics[width=.45\textwidth,height=0.3\textwidth]{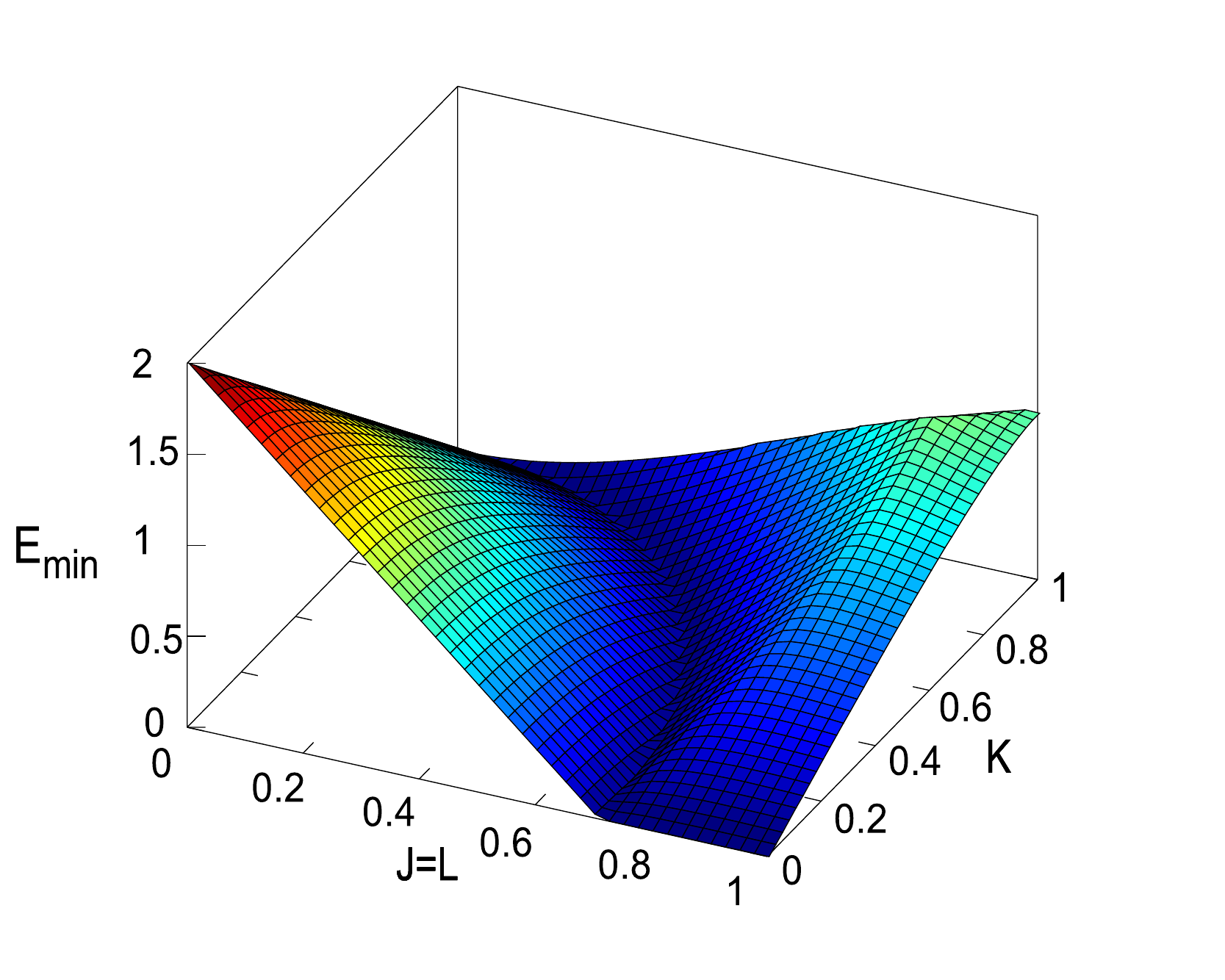}
       \caption{(Color online) The minimum energy gap of the vortex free sector with $Z=1$. The critical point can be observed along the $\sqrt{J^2 + K^2 + L^2}=1$ line. The system is gapless when $J=L>1/\sqrt{2}$ and $K=0$. More generally if any of the parameters $J$, $K$, or $L$ are zero the system is gapless beyond the phase transition. }
       \label{fig:Emin}
\end{figure}
\begin{figure}
\includegraphics[width=.4\textwidth,height=0.35\textwidth]{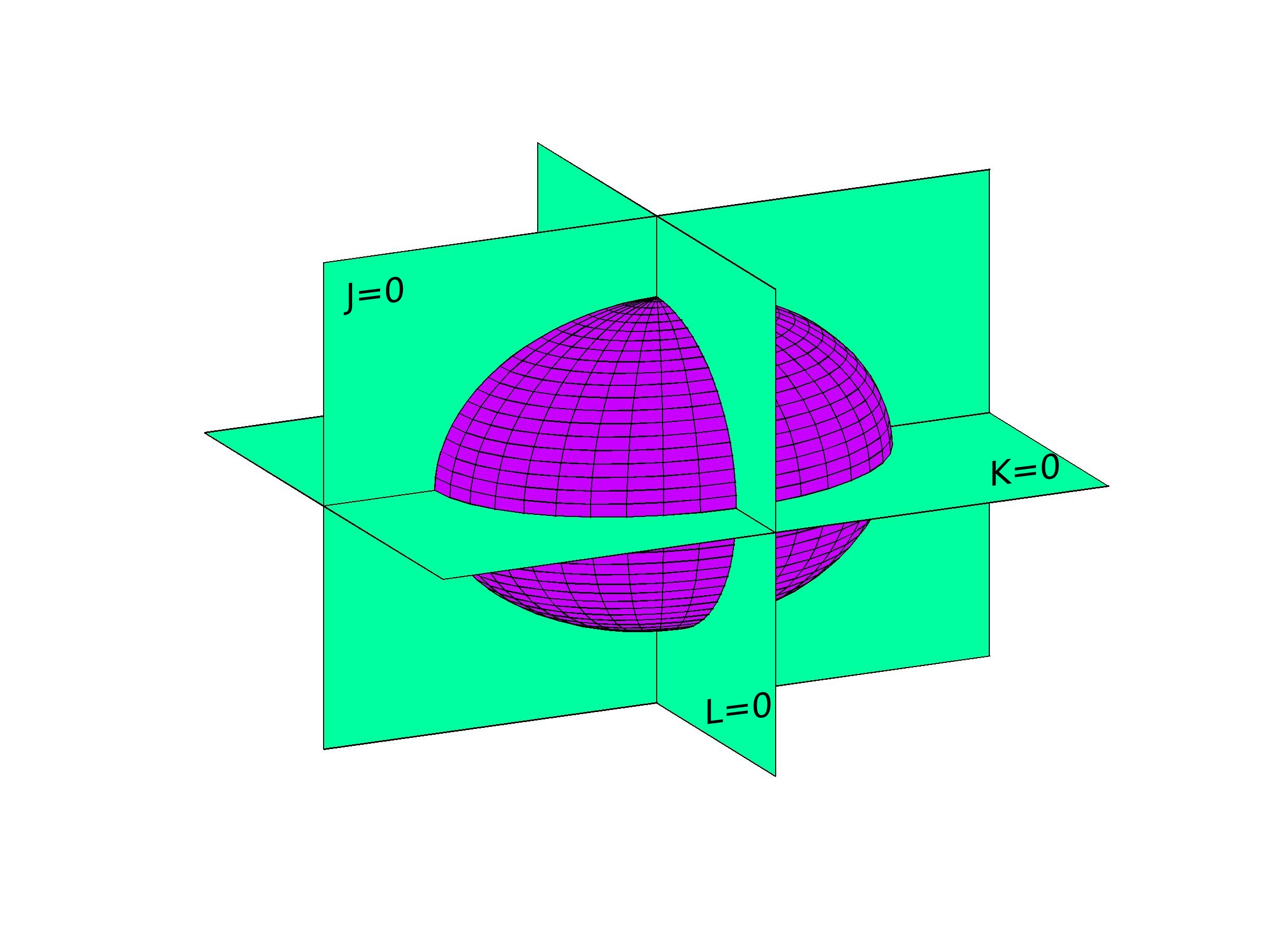}
       \caption{(Color online) Schematic of the system phase diagram. The surface of the sphere of radius $Z$ indicates the critical boundary between abelian and non-abelian phases. Inside the sphere we have a gapped abelian phase. Outside the sphere we are in a gapped non-abelian phase, provided we are not on the $J=0$, $K=0$ or $L=0$ planes indicated in light green. On these planes the system is gapless.}
       \label{fig:Phase}
\end{figure}
\begin{figure*}[ht]
\centering
\subfigure[$Z=1$,$J=K=L=0.5$]{
\includegraphics[width=.3\textwidth,height=0.25\textwidth]{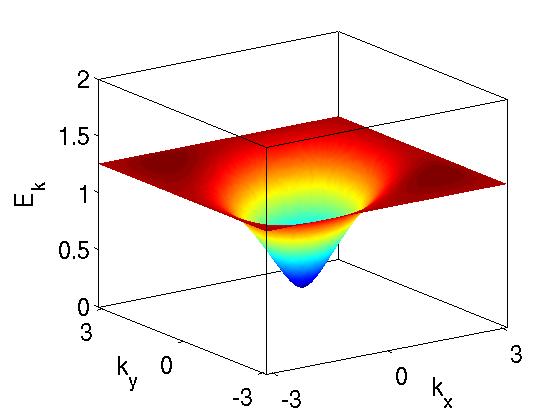}
\label{fig:abelian}
}
\subfigure[$Z=1$,$J=K=L=1/\sqrt(3)$]{
\includegraphics[width=.3\textwidth,height=0.25\textwidth]{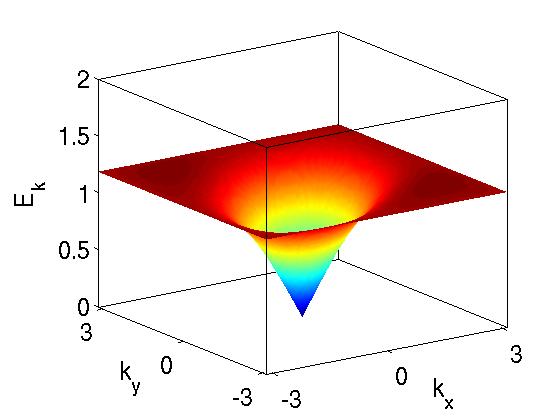}
\label{fig:critical}
}
\subfigure[$Z=1$,$J=K=L=0.6$]{
\includegraphics[width=.3\textwidth,height=0.25\textwidth]{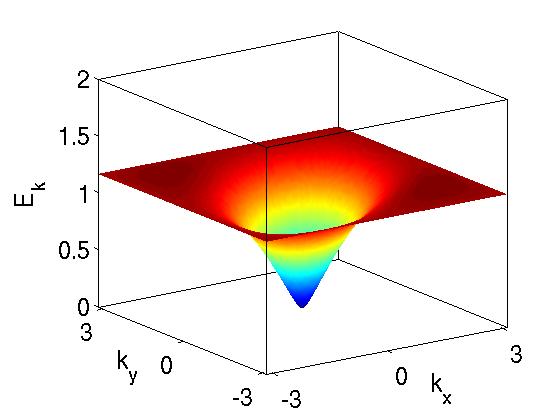}
\label{fig:non-abelian}
}
\caption[]{(Color online) Dispersion Relation $E_\bk$ for (a) the abelian phase, (b) the phase transition (c) the non-abelian phase.}
\label{fig:VF} 
\end{figure*}

In position space the eigenvalues of the BdG equation come in positive and negative pairs with eigenvectors of the form $(U,V)^T$ and $(V^*, U^*)^T$ respectively. In momentum representation however the positive and negative eigenvectors take the form $(U,V)^T$ and $(-P V^*, P U^*)^T$ respectively where the $3 \times 3$ diagonal matrix $P$ has elements $P_{11}=1$, $P_{22}=e^{-i k_x}$, and $P_{33}=e^{-ik_y}$.  

From $(\ref{eq:HfK})$ we see that the effect of the time reversal operator $T$ is to change the signs of the elements $(2,3)$ and $(3,2)$ in the matrices above. It is then straightforward to see that $ H(\bk) = (T H(-\bk))^*$. Fermionic eigenvectors in the vortex free sector are thus related to their time-reversed counterparts by conjugation and the reversal of momenta. However we again point out that these eigenvectors represent fermionic creation and annihilation operators from time reversed sectors with different toric code vacuum states.   
    
The doubled $(\pm E)$ symmetry of the spectrum means that the eigenvalue equation can be written as a cubic polynomial equation and analytical expressions for the eigenvalues can be obtained, see the appendix.  However to calculate the geometry of the phase boundary it is more straightforward to observe that within the Abelian phase the minimum gap always occurs  at $\bk  = 0 $. As this coincides with when $\Delta_{nm} = 0$ one can calculate exactly where the phase transition lies by a straightforward diagonalization of the $\xi$ matrix above with $\bk=0$. The eigenvalues are calculated to be
\be
E= \pm 2 (Z + a \sqrt{J^2 +K^2+L^2})
\label{eq:EV}
\ee
where $a=-1$, $0$ or $1$ and therefore the minimum energy gap is given by $|2 Z - 2 \sqrt{J^2 +K^2 + L^2}|$. The phase transition thus occurs at $Z = \sqrt{J^2 + K^2 + L^2}$.  This is verified in Figure \ref{fig:Emin} where the minimum energy gap as a function of $J=L$ and $K$ was calculated by numerical diagonalization. 

That the minimum energy is obtained at $\bk=0$ also holds true along the $J=K=L$ line so long as $J=K=L < \frac{\sqrt(3)}{2} Z $. We see therefore that the gap closes and opens linearly as a function of  $J$ along this line.  The dispersion relations for the A phase, the critical point , and the gapped B-phase along the $J=K=L$ line are shown in Figure \ref{fig:VF}.

\begin{figure}[ht]
\centering
\subfigure[$Z=1$,$J=L=1/\sqrt{2}+0.2$, $K=0$.]{
\includegraphics[width=.3\textwidth,height=0.2\textwidth]{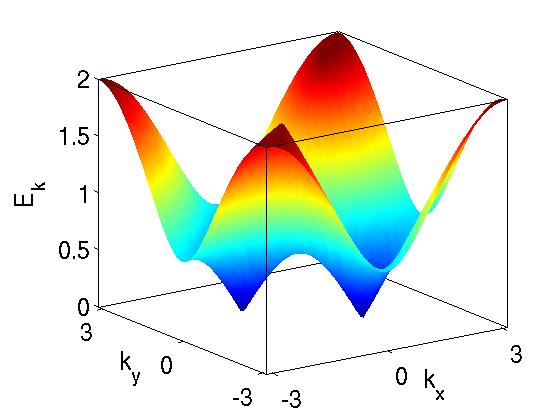}
\label{fig:KZ}
}
\subfigure[$Z=1$,$J=L=1/\sqrt{2}+0.2$, $K=0.2$]{
\includegraphics[width=.3\textwidth,height=0.2\textwidth]{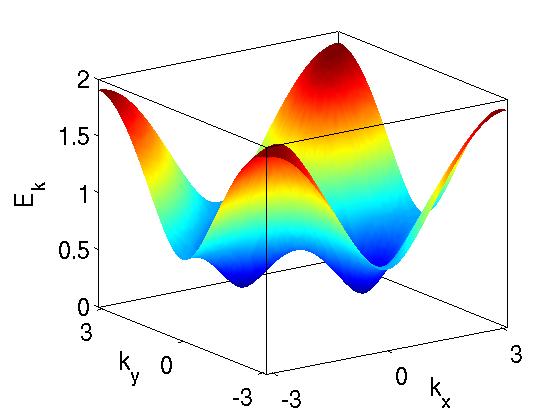}
\label{fig:KNZ}
}
\caption[]{(Color online) Dispersion Relation $E_\bk$ for (a) the gapless B-phase with $K=0$, (b) the gapped B-phase with $K=0.2$.}
\label{fig:VF_Ksmall}
\end{figure}

It is not generally true however that the minimum energy occurs when $\bk=0$. This is perhaps most striking when one of the parameters $J$, $K$, or $L$ are zero, see the appendix and Figures \ref{fig:Emin} and \ref{fig:Phase}. On these planes and outside the radius of the sphere we are in a gapless B-phase. In order to open a gap we must move off these planes.  If for example, $K=0$ this gapless phase occurs because of two Dirac cones at $\pm \bk $. Letting $K>0$ but with $K << J=L$ opens a mass gap in much the same way as the three-body terms do so in the original honeycomb model, see Figure \ref{fig:Emin} and Ref. \onlinecite{Kitaev06}. In Figure \ref{fig:VF_Ksmall} we also plot the dispersion relations showing the Dirac cones and the opening of the gap with $K>0$.  Calculations around the $L=0$ or $J=0$ planes reveal similar properties. 

\section{Ground state degeneracy on a torus}

The ground state degeneracy on a torus first observed by Ref. \onlinecite{YaoKivelson07} may be calculated within the Hartree-Fock-Bogoliubov formalism in the following way.  We have already fixed the vacua to be the toric-code states so we know by the Bloch-Messiah-Zumino theorem \cite{BlochMessiah62,Zumino62} that we can write
\be 
\ket{\text{gs}} = \prod_{m=1}^p a^\dagger_m \prod_{l\ne m} (u_{l} + v_{l} a^\dagger_{l} a^\dagger_{\bar{l}} )\ket{\{\varhexagon, \triangledown,\blacktriangle\}, \{\emptyset\} }. 
\label{eq:BCS_YK}
\ee
where the $a^\dagger$'s are the canonical fermionic raising operators gotten from the $c^\dagger$ by performing a singular value decomposition on the $U$ sub-matrix of the full eigenvector matrix $\mathcal{W}$, see for example Ref. \onlinecite{Kells09b}. The number $p$ gives the number of fully occupied un-paired fermions in the ground state and dictates the fermion number parity. In the extreme case where all fermions are unpaired the wavefunction  $\prod _{m=1}^p a^\dagger_m \ket{\text{vac}}$ is the completely filled Fermi sea.

On a torus our vacuum state takes the form $\ket{\{\varhexagon, \triangledown,\blacktriangle \},l^{(x)}_0 l^{(y)}_0 , \{\emptyset\}}$ where $l^{(x)}_0$ and  $l^{y}_0$ are the eigenvalues of the operators $L^{(x)}_0$ and $L^{(y)}_0$ defined above.  For the vortex free sector on a torus we can use the Bloch Hamiltonian (\ref{eq:Hm}) to calculate energies, eigenvectors and thus the values of $u_{l}$ and $v_{l}$.  In this case the allowed values of $k_\alpha$ in the various homology sectors on the torus of size $(N_x,N_y)$ are $\theta_\alpha + 2 \pi \frac{n_\alpha}{N_\alpha}$ for integer $n_\alpha=0,1,...N_\alpha-1$, where the four topological sectors, $(l^{(x)}_0,l^{(y)}_0)=(\pm 1,\pm 1)$ have values of $\theta_{\alpha}$ given by $\theta_\alpha = (\frac{l_0^{(\alpha)}+1}{2})\frac{\pi}{N_\alpha} $.  For a large torus one expects that these discretized $\bk$-values becomes so close as to give approximately the same ground state energy for each homology sector i.e.:
\be
E_{\text{gs}}= - \sum_\bk \sum_{n=1}^3 E_{\bk,n}/2.
\ee
This value comes directly from the assumption that all values $E_{\bk,n}$ associated with the $\gamma^\dagger$ operators are positive. However, cases where some of the $\gamma^\dagger$ must be associated with negative energy solutions do occur and in these cases the ground state energy is raised. This is precisely what happens in the $(-1,-1)$ homology sector.  In the non-abelian phase of this system any arrangement of the $\mathcal{W}$ matrix such that  positive eigenvalues (\ref{eq:EV}) at $\bk=0$ are associated with $\gamma^\dagger$ operators ensures that the matrix $U$ (i.e the upper left quadrant of $\mathcal{W}$) has one column that has just zeros. It is therefore rank deficient and, by the Bloch-Messiah-Zumino theorem, the ground state has one fully occupied mode (i.e with $u=0$ and $|v|=1$) with momentum $\bk=0$. 

Assuming we are working on an even-even lattice, the possibility of having odd fermion number parity is excluded by the fact that we are in the vortex-free sector. This means we must switch columns $(U_{l},V_{l}) \leftrightarrow (- P V_{l}^*,P U_{l}^*)$ of the eigenvector matrix describing the annihilation and creation of Bogoliubov fermions and perform the singular value decomposition of the new $U$ matrix. The switching of columns of the matrix effectively changes an occupied mode for an empty one thereby giving the correct fermionic parity number. However, it also raises the ground state energy in this sector to $E_{\text{gs}}=- \sum_{\bk ,n} E_{\bk,n}/2 +E_{\bs{0},1}$. If the system is gapped then the ground state of this sector is higher than that of the other three fully or partially anti-periodic sectors, leading to a reduction in the topological degeneracy.

\section{Conclusion and Overview} 

We have shown that the star-lattice Kitaev model may be mapped to a system of fermions hopping on an effective Kagome lattice with a $\ZZ_2$ gauge field. The fermionic vacua are explicitly shown to be toric code states of the effective Kagome spin lattice. The abelian phase of the model is inherited from the fermionic vacua and time reversal symmetry is broken at this level. We have shown that, as in the original honeycomb system, there are three distinct phases: a gapped Abelian phase, a gapless B-phase and a gapped non-Abelian B-phase.  The boundary between the phases can be understood as a sphere of radius $Z$ in the parameter space of the coupling strengths $J,K$ and $L$ around the triangles. The gapped Abelian phase lies inside the sphere. The gapless B-phase lies along the trivial planes $J=0$, $K=0$ or $L=0$  outside the sphere. These instances correspond to the cases where the underlying lattice is bipartite. Off these planes and outside the sphere the system is in a gapped non-Abelian phase with spectral properties similar to those of the gapped non-abelian phase of the original honeycomb lattice model. As in the original system the ground state degeneracy on a torus can be determined by using the Bloch-Messiah theorem. 

\section{Acknowledgements}
The authors would like to thank Charles Nash, Steve Simon, R. W. D. Nickalls, Elias Tsigaridas and Ioannis Emiris for interesting discussions. This work has been supported by Science Foundation Ireland through the President of Ireland Young Researcher Award 05/YI2/I680, the Principal Investigator Award 08/IN.1/I1961 and the Research Frontiers Programme Award 08/RFP/PHY1462. 


\begin{thebibliography}{15}
\bibitem{TRsym} Note that in any odd size system, which can include certain instances of the original honeycomb lattice model,  time reversal reversal symmetry is explicitly broken. The two fold degeneracy must also be present in these cases \cite{Chen07}.
\end{thebibliography}

\section{Appendix}
In this appendix we will derive the explicit dispersion relations for the ground state sectors as well as the exact locations of the Dirac cones that occur in the gapless B-phase. The characteristic equation of the $6\times 6$ matrix in (\ref{eq:Hk}) is
\begin{equation}
\lambda^{6} + a \lambda^{4} + b \lambda^{2} + c = 0,\label{eq:characteristic_poly}
\end{equation}
where
\begin{eqnarray}
a &=& -8 R({\bf 0}) - 12 Z^{2}  \\
b &=& 16 ((R({\bf 0}) + Z^{2})^{2} - 2 Z^{2}(R(\bk)- Z^2)) \nonumber \\
c &=& -64 Z^2 (J^{4} + K^{4} + L^{4} + Z^{4} + 2(T(\bk) - Z^{2} R(\bk))) \nonumber
\end{eqnarray}
with $R(\bk) = J^2 \cos (k_{x})  + K^2 \cos (k_{x}- k_{y}) + L^2 \cos (k_{y})$ and $T(\bk) = K^{2}L^{2}\cos(k_{x}) + J^{2}L^{2} \cos (k_{x} - k_{y}) + J^{2} K^{2} \cos (k_{y})$.

Equation (\ref{eq:characteristic_poly}) can be reduced to a degree three polynomial by simply substituting $\lambda^{2} = x$. A third degree univariate polynomial equation can be solved exactly and the solutions can be given as follows \cite{CardanoWitmer, Nickalls93}. Let $q$ and $r$ be defined as
\begin{eqnarray}
q &=& \frac{3 b - a^{2}}{9},\nonumber \\
r &=& \frac{9 a b - 27 c -2 a^3}{54},
\end{eqnarray}
and the discriminant of this polynomial be $p = q^{3} + r^{2}$. Also, taking $s$ and $t$ as
\begin{equation}
s = \sqrt[3]{r + \sqrt{p}},\; \mbox{and} \;
t = \sqrt[3]{r - \sqrt{p}},
\end{equation}
we get three roots $x_{1}, x_{2}, x_{3}$ as
\begin{eqnarray}
x_{1} &=& s + t - \frac{a}{3} \nonumber \\
x_{2} &=& -\frac{1}{2}(s + t) - \frac{a}{3} + \frac{\sqrt{3}}{2} (s-t) i\nonumber \\
x_{3} &=& -\frac{1}{2}(s + t) - \frac{a}{3} - \frac{\sqrt{3}}{2} (s-t) i.
\end{eqnarray}
Now, for $p > 0$, two of these are complex conjugate roots. However, as all the eigenvalues are known to be real we may disregard this case. For $p \leq 0$, we let $\theta = \arccos ({\frac{r}{\sqrt{-q^{3}}}})$, and the three real roots can be given as
\begin{eqnarray}
x_{1} &=& 2 \sqrt{-q} \; \cos (\frac{\theta}{3} ) - \frac{a}{3} \nonumber \\
x_{2} &=& 2 \sqrt{-q}\; \cos (\frac{\theta + 2\pi}{3}) - \frac{a}{3} \nonumber \\
x_{3} &=& 2 \sqrt{-q} \; \cos (\frac{\theta + 4\pi}{3}) - \frac{a}{3}.
\end{eqnarray}
The real roots of a generic cubic univariate polynomial equation have recently been classified in terms of its coefficients using real algebraic geometry techniques \cite{Wies94, EmiTsi04}.  For $p = 0$, $r = \pm \sqrt{-q^{3}}$ and so $x_{2} = x_{3}$.  For all the cases we have observed the dispersion relation (the minimum energy band) is given by $E_{\bk}=\lambda_2=\sqrt{x_2}$.

Gapless phases in the system may be identified by setting $c=0$ or equivalently setting $\det(H)=0$. From this we obtain a quadratic equation in the variable $z=Z^2$:
\be
z^2+2 R(\bk) z + 2 T(\bk) +J^4+K^4+L^4 =0.
\ee
with generic solutions
\be
Z= \sqrt{R(\bk) \pm \sqrt{R(\bk)^2- [2 T(\bk)+J^4+K^4+L^4]}}.
\ee

If $\bk=0$ we have $ R(0)^2= [2 T(0)+J^4+K^4+L^4]$ and the quantum phase transition can be seen to occur on the surface of the 3-sphere $Z=\sqrt{R(0)}$ in the space of the coupling parameters $J$,$K$ and $L$, see Figure \ref{fig:Phase}. 

Other interesting cases occur when either of the parameters $J$,$K$ or $L$ vanish. In these situations we have
\bea 
R(\bk)^2 &-& [2 T(\bk)+J^4+K^4+L^4 ]  \\ \non
&=& i [K^2 \sin(k_x-k_y) - L^2 \sin(k_y)],  \quad J=0  \\ \non
&=& i [J^2 \sin(k_x) +L^2 \sin(k_y)],      \; \quad \quad \quad K=0  \\ \non
&=& i [J^2 \sin(k_x) +K^2 \sin(k_x-k_y)],   \quad L=0   
\eea
which, because the parameter $Z$ is real, must vanish. This can be used to set conditions on parameters $k_x$ and $k_y$. In the cases above if the two non-zero terms are set to be equal (for example $L=K=M,J=0$), the three equations above reduce to (i) $k_x=2k_y$ (ii) $k_x=-k_y$ (iii) $2k_x=k_y$  respectively. Substitution into the remaining part $Z=\sqrt{R(\bk)}$ gives the zero energy solutions at 
\bea
(k_x,k_y) &=& \pm (2l,l) \;\quad J=0, L=K=M \non \\ 
&& \pm (l,-l) \quad K=0, J=L=M \non \\
&& \pm(l,2l) \;\quad L=0, J=K=M \non
\eea
where $ l = \cos^{-1}(Z^2/2M^2)$. If $M<1/\sqrt{2}$ we see that $l$ is complex but as we have already shown these values of $J,K$ and $L$ corresponds to the gapped phase inside the sphere $Z=\sqrt{R(\bs{0})}$. The zero energy solutions described here form the sharp points of the Dirac cones at these values of $k_x$ and $k_y$. A gap may be opened in the spectrum at these conical points by letting each of the parameters  $J$, $K$ and $L$ have non-vanishing values such that the condition $\sqrt{R(\bs{0})}>Z$ is fulfilled.

\end{document}